\documentclass[aps,prl,floatfix,amsmath,amssymb,twocolumn,showpacs]{revtex4-1}
\usepackage{amsmath,amssymb,graphics, mathrsfs,bm}
\usepackage{pstricks}
\usepackage{graphicx}
\usepackage[pdftex, bookmarks={false}, pdfauthor={Rudro Rana Biswas}, pdftitle={Majorana fermions in vortex lattices}]{hyperref}
\usepackage{subfigure}
\usepackage{parskip}
\allowdisplaybreaks[1]

\newcommand\ba{\begin{array}}
\newcommand\ea{\end{array}}
\newcommand\nn{\nonumber}

\newcommand\ri{\right}
\renewcommand\le{\left}

\newcommand{\feyn}[1]{#1\kern-0.45em/}

\renewcommand\a{\alpha}

\renewcommand\b{\beta}

\renewcommand\c{\psi}

\renewcommand\d{\delta}
\newcommand\mbd{\mbs{\delta}}

\newcommand\D{\Delta}

\newcommand\f{\phi}

\newcommand\F{\Phi}

\newcommand\g{\gamma}

\renewcommand\k{\kappa}

\renewcommand\l{\lambda}

\newcommand\n{\nu}

\newcommand\p{\pi}
\newcommand\mbp{\mbs{p}}

\newcommand\mbq{\mbs{q}}

\newcommand\rr{\rho}

\newcommand\mbr{\mbs{r}}
\newcommand\mbR{\mbs{R}}

\renewcommand\th{\theta}
\newcommand\vth{\vartheta}

\newcommand\Th{\Theta}

\newcommand\x{\xi}

\newcommand\pd{\partial}
\newcommand\mc{\mathcal}

\newcommand\mbs{\boldsymbol}

\newcommand\msf{\mathsf}

\begin{document}
\title{Majorana fermions in vortex lattices}
\author{Rudro R. Biswas}
\email{rrbiswas@illinois.edu}
\affiliation{University of Illinois at Urbana-Champaign, 1110 W.\ Green St., Urbana, IL 61801}

\begin{abstract}
We consider Majorana fermions tunneling among an array of vortices in a 2D chiral p-wave superconductor or equivalent material. The amplitude for Majorana fermions to tunnel between a pair of vortices is found to necessarily depend on the background superconducting phase profile; it is found to be proportional to the sine of half the difference between the phases at the two vortices. Using this result we study tight-binding models of Majorana fermions in vortices arranged in triangular or square lattices. In both cases we find that the aforementioned phase-tunneling relationship leads to the creation of superlattices where the Majorana fermions form macroscopically degenerate localizable flat bands at zero energy, in addition to other dispersive bands. This finding suggests that tunneling processes in these vortex arrays do not change the energies of a finite fraction of Majorana fermions, contrary to previous expectation. The presence of flat Majorana bands, and hence less-than-expected decoherence in these vortex arrays, bodes well for the prospects of topological quantum computation with large numbers of Majorana states.
\end{abstract}
\maketitle

The last two decades have seen a flurry of activity in the fields of topological states of matter and topological quantum computation(TQC)\cite{2008-nayak-fk}. One candidate particle for TQC with the requisite exotic `anyonic' characteristics is the Majorana fermion state in the vortex of a chiral p-wave superconductor\cite{2001-ivanov-fk,2004-stern-fk}. Such Majorana states are predicted to also arise in the $\n=5/2$ fractional quantum Hall state\cite{2000-read-fk}, Sr$_{2}$RuO$_{4}$\cite{1995-rice-uq}, the topological insulator surface state\cite{2008-fu-yq} and in nanowires with spin-orbit coupling\cite{2010-sau-fk}, to name but a few widely-reported possibilities. In 2D systems there have been proposals to create multiple vortices in the host superconductor and to perform TQC by `braiding' them, i.e, by physically moving the vortices around each other\cite{2001-ivanov-fk}. Practical implementations of TQC would require bringing together a large number of Majorana fermions and much attention has been paid to the splitting of their energy levels due to inter-vortex tunneling. These level splittings would introduce decoherence on timescales inversely proportional to their magnitudes and place duration constraints on computation processes. In 2D systems, prospects for realizing large well-controlled vortex arrays are bright\cite{2012-eley-fk}, potentially enabling the creation of vortex lattices with Majorana fermions, where TQC by `braiding' the vortices is possible. Several works have addressed properties of inter-vortex Majorana tunneling both in the case of a single vortex pair\cite{2009-cheng-uq,2009-kraus-fk} as well as in vortex lattices\cite{2000-franz-uq, 2001-vafek-fk, 2006-grosfeld-fk}. In particular, for infinite vortex lattices with finite magnetic penetration depth, $\l$, methods exist for accurately computing the quasiparticle spectrum\cite{2000-franz-uq, 2001-vafek-fk}. In this paper we shall consider the opposite limit where $\l$ is much larger than the lattice size. This condition can be satisfied for 2D SC films and is also the limit where TQC with Majorana fermions is best understood.

\begin{figure}[ht]
\begin{center}
\subfigure[]
{\resizebox{7.3cm}{!}{\includegraphics{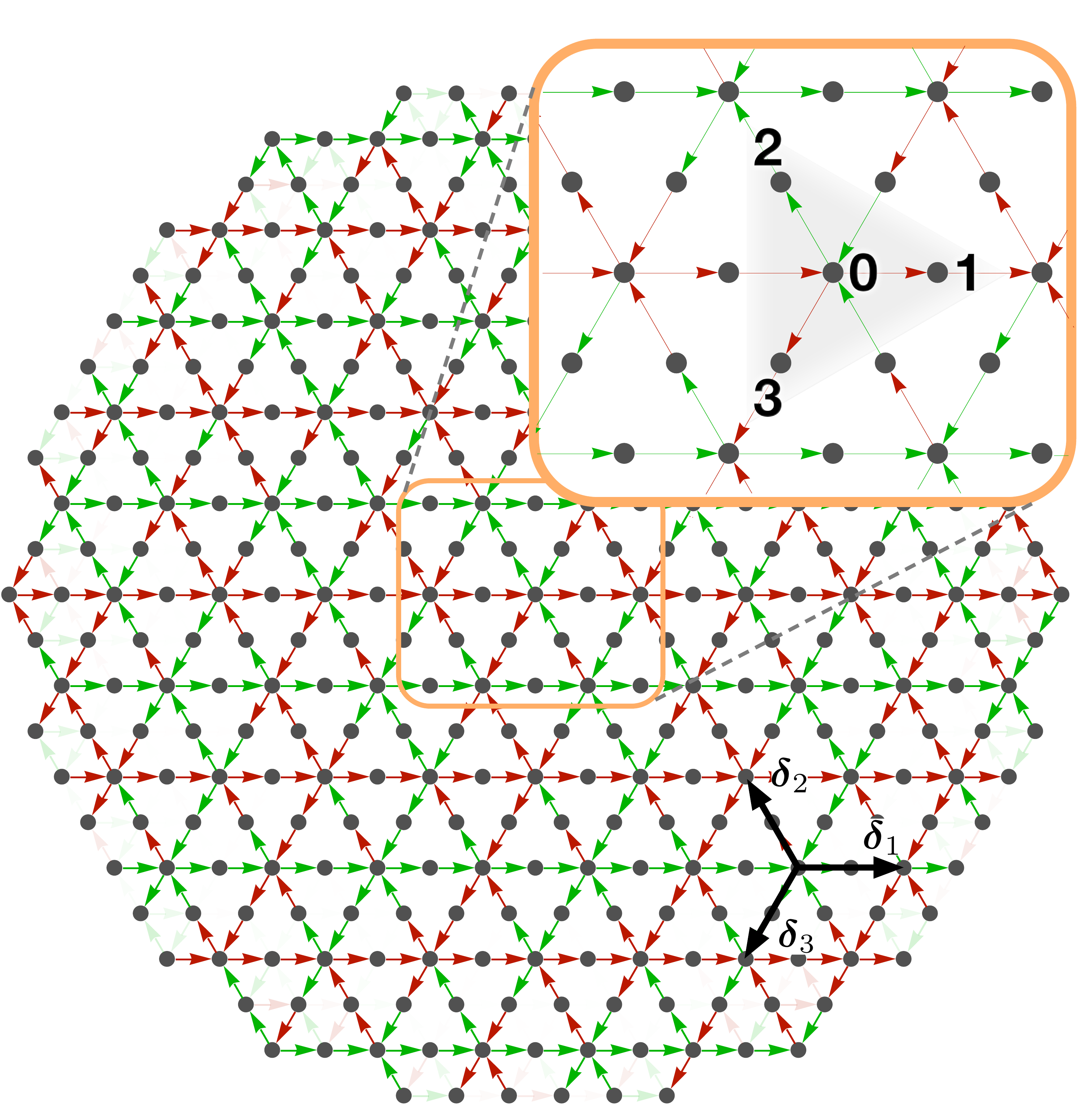}}
\label{fig-fulltriangular}}
\subfigure[]
{\resizebox{7.3cm}{!}{\includegraphics{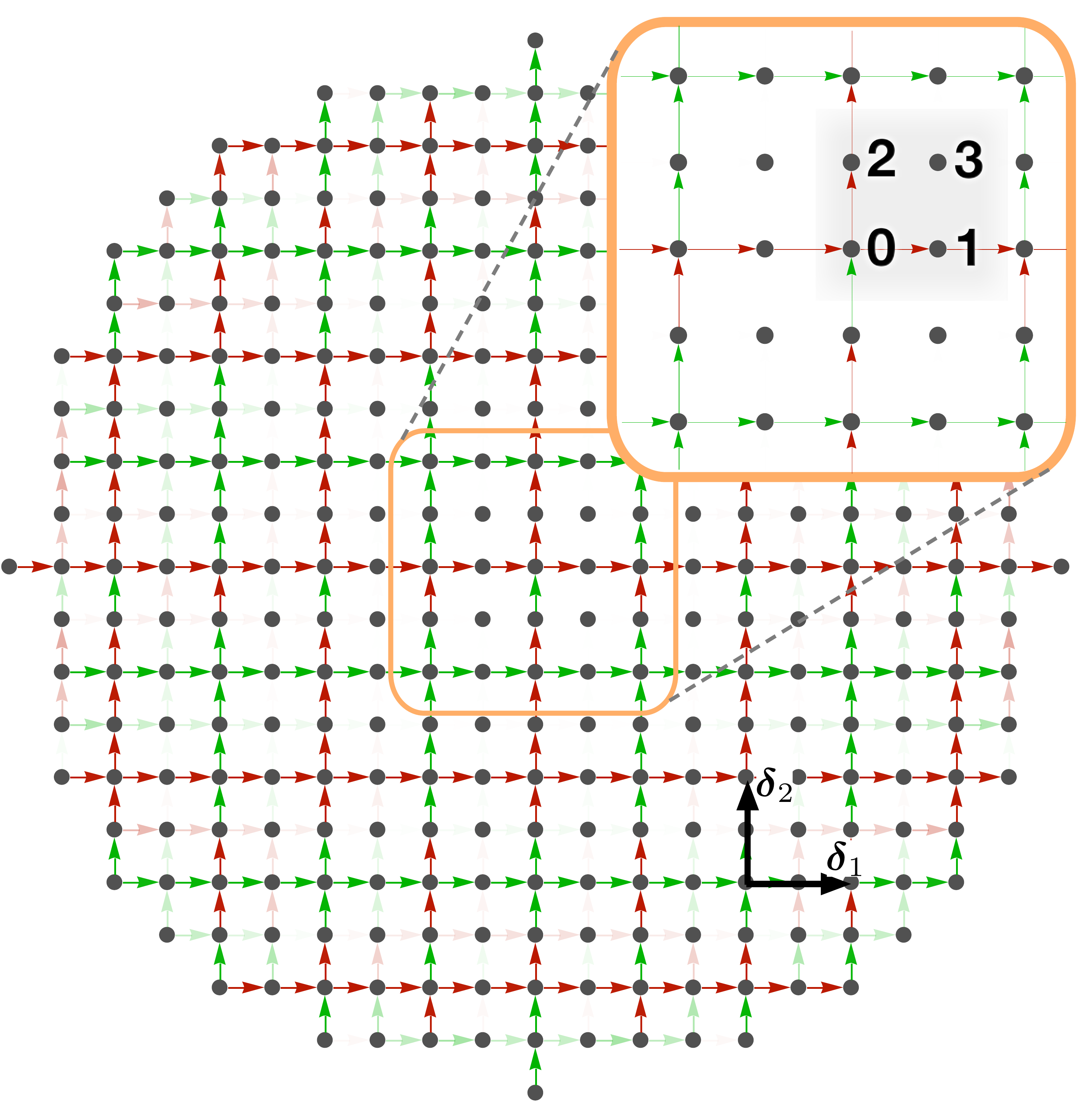}}
\label{fig-fullsquare}}
\caption{(Color online) Triangular and square vortex lattices (vortices are the black dots) with circular boundaries showing the magnitudes and signs of the purely imaginary nearest neighbor hopping amplitudes between Majorana states, encoded by the opacity and color (green/red=$+/-$) of the directed bonds respectively. The \emph{translation-invariant $Z_{2}$ flux} per plaquette is negative/positive for the triangular/square lattice as there are an odd/even number of negative (red) directed bonds around each loop. Inset are zoomed sections showing the numbered four-site bases of the superlattices.}
\label{fig-fulllattice}
\end{center}
\end{figure}

In this work we analyze the problem of Majorana fermions tunneling between vortices in a chiral p-wave superconductor (or its equivalents\cite{2008-fu-yq}) and discover that the order parameter phase plays a crucial role, leading to surprising and promising features. We find that the purely imaginary tunneling amplitude $t_{AB}$ between the Majorana fermions at two vortices $A$ and $B$ depends on the order parameter phases $\a$ and $\b$ at the two vortices according to $t_{AB}\propto i \sin[(\a-\b)/2]$. This dependence is a feature of Majorana fermions since they require $t_{AB}=-t_{BA}$ (see discussion following Eq.~\eqref{eq-hoppingampl}). The amplitude also decays exponentially and oscillates with the distance between the vortices, in agreement with previous work with two vortices\cite{2001-kitaev-fk,2009-cheng-uq,2009-kraus-fk}. Numerically computing nearest neighbor hopping amplitudes in a large triangular/square lattice of vortices shows that in the central regions of these lattices and in particular, covering almost the entire lattice when it has a circular boundary, we find a superlattice emerging due to amplitudes vanishing on certain bonds. The Hamiltonian there is found to be translation-invariant with a superlattice lattice constant that is twice that of the original vortex lattice, up to appropriate $Z_{2}$ gauge transformations. In both the square and triangular lattice cases we find \emph{localizable flat bands} of Majorana fermions at zero energy, i.e, with no energy shift as compared with their energy in an isolated vortex. This \emph{macroscopic degeneracy of the ground state} should dominate thermodynamic and transport characteristics of the Majorana fermions in these lattices and will also result in stronger quantum coherence than has been believed till now. We also find bands with non-zero energy which are gapped in the case of the triangular lattice and gapless for the square lattice.

\textbf{Inter-vortex tunneling amplitude:} We consider a network of vortices centered at locations $\le\{\mbr_{k}\ri\}$ with the same vorticity $\ell$. Following past research\cite{2008-nayak-fk}, we shall neglect the effects of the magnetic vector potential and assume that $\l$ is large compared to the lattice size. The chiral p-wave superconducting(SC) order parameter is given by $\D(\mbr)(\pd_{x}+i\pd_{y})$, where $\D(\mbr) = f(\mbr)e^{i\f(\mbr)}$. The SC amplitude $f(\mbr)$ is real, positive and equal to $\D_{0}$ at all places other than within a distance of $\sim \x$ (coherence length) from the vortex cores, where it drops to zero with a functional dependence $f_{0}(|\mbr-\mbr_{k}|)$ near the $k^{th}$ core. $\f(\mbr)$ is the phase of the SC order parameter. We now isolate two vortices $A$ and $B$, with vortex $B$ located at a polar coordinate $(\mc{R},\Th)$ with respect to vortex $A$. The polar coordinates of a generic point is denoted by $(r,\th)$ or $(\rr, \vth)$ with respect to $A$ or $B$ respectively. The (real) phase $\f(\mbr)$ takes the value $\a, \b$ immediately to the `right' of the vortices $A$ and $B$ respectively, so that near $A$ and $B$ the vortex phase $\f(\mbr)$ varies as $\a + \ell\th$ and $\b+\ell\vth$ respectively.

The unperturbed zero energy Majorana wavefunctions near each of the two vortices, that exist only if the vorticity $\ell$ is odd, are given by (for a derivation see \cite{2010-cheng-fk} for the $\ell=1$ case)
\begin{subequations}\label{eq-trialwavefunctions}
\begin{align}
\c_{A} &\propto \le(\ba{c} e^{i\a/2}e^{i(\ell+1)\th/2} \\ -e^{-i\a/2}e^{-i(\ell+1)\th/2} \ea\ri) \F_{\ell}(r),\\
\c_{B} &\propto \le(\ba{c} e^{i\b/2}e^{i(\ell+1)\vth/2} \\ -e^{-i\b/2}e^{-i(\ell+1)\vth/2} \ea\ri) \F_{\ell}(\rr),
\end{align}
\end{subequations}
where
\begin{align}
\F_{\ell}(r) = J_{(\ell+1)/2}(\k r)\exp\le(-\frac{1}{v_{F}}\int_{0}^{r}f_{0}(r')dr'\ri)
\end{align}
and $\k = k_{F} \sqrt{1-\D_{0}^{2}/(v_{F}k_{F})^{2}}$. We shall assume that the lowest energy states in the presence of tunneling between the two vortices is spanned by these two wavefunctions.

Introducing the notation $\msf{O}_{\msf{\D}} = \le\{\D,(\pd_{x} + i \pd_{y})\ri\}/2$ and $\msf{H}$ the single-particle part of the effective Hamiltonian, the amplitude for tunneling from vortex $A$ to $B$ is given by:

\begin{align}
t_{AB} &\propto \int d^{2}x\,\c_{B}^{\dag}\cdot\le(\ba{cc} \msf{H} & \msf{O}_{\msf{\D}}\\-\msf{O}_{\msf{\D}}^{*} &- \msf{H}^{*}  \ea \ri)\cdot\c_{A} \nn\\
&= \int d^{2}x\,\bigg[\c_{B}^{\dag}\cdot\overbrace{\le(\ba{cc} \msf{H} & \msf{O}_{\msf{\D_{A}}}\\-\msf{O}_{\msf{\D_{A}}}^{*} &- \msf{H}^{*}  \ea \ri)\cdot\c_{A}}^{0} \nn\\
&\quad+ \c_{B}^{\dag}\cdot\le(\ba{cc} 0 & \msf{O}_{\msf{\D}} - \msf{O}_{\msf{\D_{A}}}\\-(\msf{O}_{\msf{\D}} - \msf{O}_{\msf{\D_{A}}})^{*} & 0 \ea \ri)\cdot\c_{A}\bigg]\nn\\
&\propto i \iint d^{2}x\, \text{Im}\big[e^{-i(\a+\b)/2}\F(\rr)e^{-i(\ell+1)\vth/2}\nn\\
&\qquad\qquad\qquad \times (\msf{O}_{\msf{\D}} - \msf{O}_{\msf{\D_{A}}})e^{-i(\ell+1)\th/2} \F(r)\big],
\end{align}
where the proportionality factors are real numbers and will be in the following equations. $\msf{\D_{A}}$ introduced in the second step above is the order parameter when only vortex $A$ is present and whose phase is given by $\a + \ell\th$, i.e, near vortex $A$, $\D\simeq \D_{A}$. Thus, the overlap integral becomes significant only if we are far enough from $A$. Meanwhile we see that the exponentially decaying factors in the functions $\F$ confine any meaningful contribution from the overlap integral to a strip of area connecting vortices $A$ and $B$ and having a width of the order of the coherence length $\x$. Finally derivatives of $\D$ near vortex $B$ will be larger by a factor of the order of $(\mc{R}/\x)$ when compared to derivatives of $\c_{A}$ and $\D_{A}$. Thus we estimate that
\begin{align}
t_{AB} &\propto i \iint_{\text{near }B} d^{2}x\, \text{Im}\big[e^{-i(\a+\b)/2}\F(\rr)\F(r)\nn\\
&\qquad\qquad\times e^{-i(\ell+1)(\vth + \th)/2} (\pd_{x} + i\pd_{y})\D(\mbr)\big].
\end{align}
Near $B$, $\D(\mbr)\approx f_{0}(\rr) e^{i\b}e^{i\ell \vth}$ and so
\begin{align}
(\pd_{x} + i\pd_{y})\D(\mbr) \approx e^{i\b}e^{i(\ell + 1)\vth}\le(f_{0}'(\rr) - \frac{\ell + 1}{\rr}f_{0}(\rr)\ri)
\end{align}
Thus,
\begin{align}
t_{AB} &\propto i\,\text{Im}\big[e^{i(\b-\a)/2} \iint_{\text{near }B} d^{2}x\, \F(\rr)\F(r)\nn\\
&\qquad \times e^{i(\ell+1)(\vth-\th)/2} \le(f_{0}'(\rr) - \frac{\ell + 1}{\rr}f_{0}(\rr)\ri)\big].
\end{align}
Since the above integrand transforms to its complex conjugate when we consider points symmetric about the line joining the two vortices, the integral is \emph{real} and so we arrive at a central result of this paper
\begin{align}\label{eq-hoppingampl}
t_{AB} &\propto i \sin\le(\frac{\b-\a}{2}\ri).
\end{align}
The hopping amplitude is purely imaginary as it must be for Majorana states. The proportionality factor involves $\F(\mc{R})$ and its derivatives which give rise to the exponential decay and oscillatory behavior that have been obtained before\cite{2009-cheng-uq}. We note here that there is an ambiguity of $2\p$ in each of the phases $\a$ and $\b$ and this leads to an ambiguity in the sign of \emph{all} amplitudes connecting a given vortex -- this is however physically irrelevant due to the $Z_{2}$ freedom in the sign of the Majorana operator.

Such dependence on the SC phase that is anti-symmetric under the exchange of the vortices is essential because if the hopping amplitude were a function only of the scalar $\mc{R}$, then it has to be zero since it has to change sign when the two Majorana states are interchanged, i.e, we require $t_{AB} = - t_{BA}$. This physics seems to have been obscured in previous research on this topic.

\textbf{Tight-binding models:} We consider nearest neighbor hopping only as by comparison all longer range hops are exponentially suppressed with distance. Numerical calculations in the following sections using the ansatz Eq.~\eqref{eq-hoppingampl} suggest simple translationally invariant superlattice models for the Majorana fermions when dealing with the triangular and square vortex lattices (similar calculations may be performed for other lattices). We consider odd vorticities $\ell$ since only in those cases can we have the Majorana states and in those cases numerical computation yields a simple picture. The SC phase profile is calculated as $\f(\mbr) = \ell \sum_{k} \arg[z(\mbr)-z(\mbr_{k})]$, where $z(x,y) = x+iy$\cite{2004-stern-fk}. While the exact realization of the hopping amplitude profile depends sensitively on the shape of the lattice boundary, in the bulk central regions of \emph{all} lattices\footnote{Examples of how the superlattice model described here arises inside the bulk of lattices with non-circular boundaries are presented in the supplemental material text and animations, hosted at \url{http://prl.aps.org/supplemental/PRL/v111/i13/e136401}.} and covering almost all of the area of a lattice with a circular boundary we find a superlattice network as shown in Figure~\ref{fig-fulllattice}. As can be seen there, many hopping amplitudes go to zero and the rest define superlattices with constant $Z_{2}$ flux per plaquette. These should thus be gauge-equivalent to translation-invariant tight binding models for the Majorana fermions, as derived below.

\begin{figure}[ht]
\begin{center}
\subfigure[DOS for triangular lattice]
{\resizebox{7cm}{!}{\includegraphics{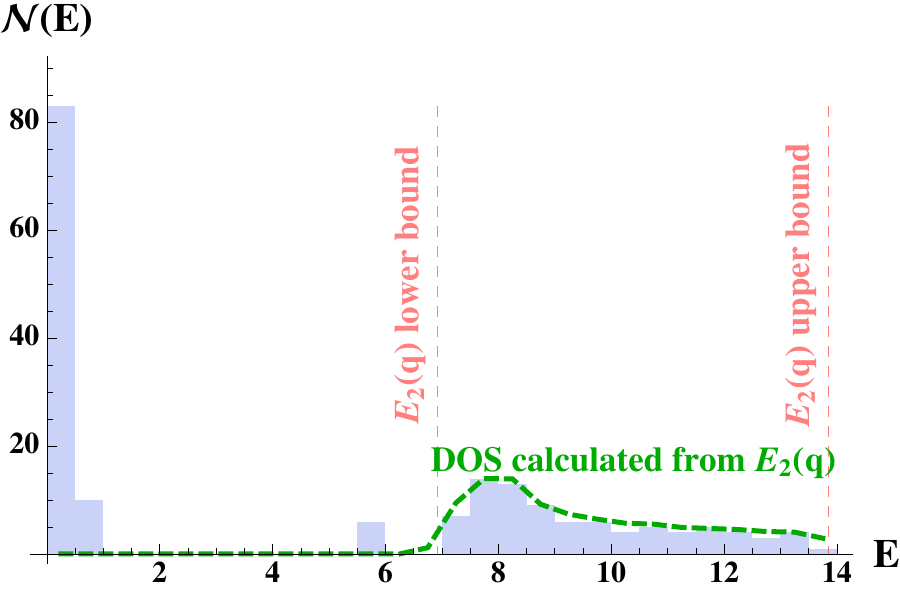}}
\label{fig-dostriangular}}
\subfigure[DOS for square lattice]
{\resizebox{7cm}{!}{\includegraphics{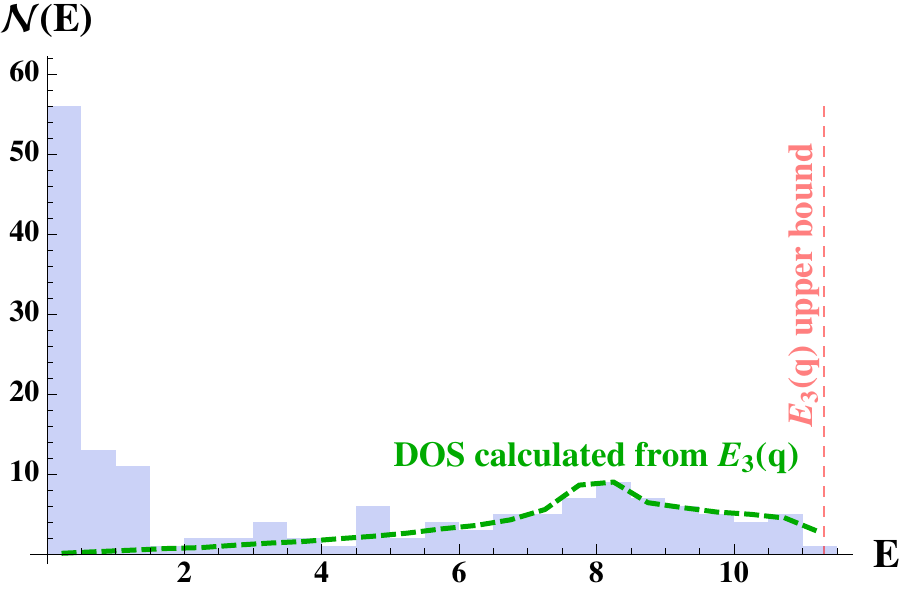}}
\label{fig-dossquare}}
\caption{(Color online) Histograms of energy eigenstates in the lattices shown in Figure~\ref{fig-fulllattice}. Both show \emph{macroscopic degeneracies at zero energy}, with some broadening due to boundary effects. The density of states (DOS) at higher energies are fit well using energy bands calculated in the main text (Eqs \eqref{eq-bandstriangular} and \eqref{eq-bandsquare}).}
\label{fig-dos}
\end{center}
\end{figure}

\textbf{The triangular lattice: } The superlattice here is a triangular lattice with a basis of 4 vortices, as shown in Figure~\ref{fig-fulltriangular}. The hopping amplitudes along the directed intra-basis bonds are the \emph{same} while the inter-site amplitudes have the opposite sign. The Hamiltonian for the bulk is thus (up to an overall scale factor):
\begin{align}\label{eq-tbmodel}
\mc{H}_{B} &= i \sum_{\mbR}\sum_{\a=1}^{3} \le(\g^{0}_{\mbR} + \g^{0}_{\mbR+\mbd_{\a}}\ri)\g^{\a}_{\mbR} + \text{h.c}.
\end{align}
The $\mbd_{\a}$ are marked on Figure~\ref{fig-fulltriangular} while $\g^{\a}_{\mbR}$ is the Majorana state operator for the $\a^{th}$ site at superlattice point $\mbR$. We choose the usual convention normalizing the $\g$ operators $\le\{\g^{\a}_{\mbR},\g^{\a'}_{\mbR'}\ri\} = 2 \d_{\a\a'}\d_{\mbR, \mbR'}$. The Hamiltonian \eqref{eq-tbmodel} can be solved by using the traveling modes\cite{2011-biswas-uq} $b^{\a}_{\mbq} = \frac{1}{\sqrt{2\mc{N}}} \sum_{\mbR} \g^{\a}_{\mbR} e^{-i \mbq\cdot\mbR}$, where $\mc{N}$ is the number of superlattice sites.
The $\mbq$ vectors span the BZ of the superlattice and the modes at $\pm\mbq$ are identified with each other: $b^{\a}_{\mbq} = \le(b^{\a}_{-\mbq}\ri)^{\dag}$ and $\le\{b^{\a}_{\mbq},b^{\b}_{\mbp}\ri\} = \d_{\a\b}\d_{\mbq,-\mbp}$. The Hamiltonian can now be rewritten as
\begin{align}
\mc{H}_{B} &= \frac{1}{2}\sum_{\mbq\in\text{BZ}}\sum_{\a,\b=0}^{3} (b^{\a}_{\mbq})^{\dag}h_{\a\b}(\mbq)b^{\b}_{\mbq},
\end{align}
where
\begin{align}
h_{\a\b}(\mbq) &= \le\{\ba{cl} 4i\le(1 + e^{-i\mbq\cdot\mbd_{\b}}\ri), & \a= 0, \b = 1,2,3,\\-4i\le(1 + e^{i\mbq\cdot\mbd_{\a}}\ri),& \a= 1,2,3,\, \b = 0,\\ 0 & \text{otherwise.} \ea\ri.
\end{align}
Finding the excitation spectrum of this model is reduced to diagonalizing $h(\mbq)$ and retaining an independent half of the particle-hole conjugate modes with non-negative eigenvalues. In this case, diagonalization yields two bands of complex fermions with energies
\begin{subequations}\label{eq-bandstriangular}
\begin{align}
E_{1}(\mbq) &= 0, \\
E_{2}(\mbq) &= 8\sqrt{1 + 2\cos\frac{\mbq\cdot\mbd_{1}}{2}\cos\frac{\mbq\cdot\mbd_{2}}{2}\cos\frac{\mbq\cdot\mbd_{3}}{2}}.
\end{align}
\end{subequations}
The eigenfunctions corresponding to the zero energy flat band $E_{1}(\mbq)$ also possess Chern number equal to zero, implying that we can form Wannier-localized Majorana states involving the bond center sites whose energies are (still) zero! As shown in Figure~\ref{fig-dostriangular}, the DOS obtained from this model is a good approximation to the numerically computed DOS of the model in Figure~\ref{fig-fulltriangular}.


\textbf{The square lattice: } The superlattice for this problem is a square lattice with a basis of 4 vortices, as shown in Figure~\ref{fig-fullsquare}. The hopping amplitudes along directed bonds can be chosen to be the \emph{same}. The Hamiltonian for the bulk is thus (up to an overall scale factor):
\begin{align}\label{eq-tbmodel2}
\mc{H}_{B} &= i \sum_{\mbR}\sum_{\a=1}^{2} \le(\g^{0}_{\mbR} - \g^{0}_{\mbR+\mbd_{\a}}\ri)\g^{\a}_{\mbR} + \text{h.c}.
\end{align}
The basis sites labeled `$3$' are disconnected from the rest of the lattice and so they form a flat band of \emph{site-localized zero energy} Majorana states! Using techniques introduced in the previous section, we find that the other three sites lead to the formation of another \emph{zero energy flat band} of Majorana fermions as well as a gapless band of complex fermions spanning the entire BZ and having the dispersion
\begin{align}\label{eq-bandsquare}
E_{3}(\mbq) = 8 \sqrt{\sin^{2}(q_{x}/2)+\sin^{2}(q_{y}/2)}.
\end{align}
As in the case of the triangular lattice, the zero energy band wavefunctions have Chern number zero and so can be used to form zero energy Wannier-localized Majorana states. As shown in Figure~\ref{fig-dossquare}, the DOS obtained from this model is a good approximation to the numerically computed DOS of the lattice in Figure~\ref{fig-fullsquare}.

For both the triangular and square lattices in Figure~\ref{fig-fulllattice}, animations scanning across all energy eigenfunctions have been provided as supplementary material.

\textbf{Discussion and conclusions:} We have discovered that the global \emph{superconducting phase profile plays an important role in the tunneling amplitudes} of Majorana fermions tunneling between vortices. This is necessary since the hopping amplitude $t_{AB}$ needs to satisfy the requirement $t_{AB} = - t_{BA}$ for Majorana fermions, which is impossible if $t_{AB}$ is a nonzero and a function only of the inter-vortex distance. This conclusion was obscured in previous research involving only two vortices, for which the relative phase $|\a-\b| = \ell \p$ is untunable. We estimate that the SC phase dependence of the tunneling amplitude manifests itself in the sinusoidal law expressed in equation~\eqref{eq-hoppingampl}. This has surprising consequences for triangular and square vortex lattices where we find emergent superlattice structures hosting \emph{zero energy localizable} Majorana fermion states that number a finite fraction of the original number of states. In the square lattice these also include single-site localized states on a quarter of all lattice sites. These zero energy states imply the existence of temporally coherent states even when the vortices are relatively close together which bodes well for the feasibility of TQC in such setups.

We emphasize here that the design and implementation of such vortex lattices is feasible\footnote{Private communication with Prof.\ Nadya Mason, UIUC.} and not just an esoteric possibility. Taking a cue from setups that have already been fabricated\cite{2012-eley-fk}, such vortex lattices may be created by depositing a patterned layer of s-wave superconducting metal like Nb or Al on a strong topological insulator(STI) such as Bi$_{2}$Se$_{3}$. The deposited layer is continuous except for holes laid out in the pattern of the required vortex lattice. Cooling this arrangement below a suitable temperature will induce superconductivity in the STI surface state that has been predicted to host Majorana states at vortices\cite{2008-fu-yq}. Turning on a magnetic field of sufficient strength will cause vortices to appear at the holes since the induced superconductor is weakest at those points, providing a strong pinning potential. Designing protocols for using these zero energy states in the vortex lattice for TQC is an open problem.

An important ingredient in obtaining the phase-dependence of the tunneling amplitude is the choice of trial wavefunctions (Eq.~\eqref{eq-trialwavefunctions}) spanning the low energy subspace, which is the same as in other previous approaches (e.g \cite{2009-cheng-uq}). These wavefunctions may be improved to take into account the mismatch between the phase of $\D(\mbr)$ and the uniform angular twisting of the SC order parameter phase, in the spirit of \cite{2000-franz-uq, 2001-vafek-fk} but modified for our scenario where the magnetic length is infinite.

Finally, while considering disorder in the tunneling amplitudes that arise due to randomness in the inter-vortex distances, our work suggests that past calculations\cite{2000-read-kx,2009-levin-ys,2009-bonderson-vn,2011-kraus-uq,2012-laumann-fk} be extended to take into account the long range effects of lattice geometry on the effective anyonic Hamiltonian, mediated by the order parameter phase profile. For e.g, in the case mentioned here the bonds with zero strength have negligible noise and do not fit into the scheme of earlier analyses. Also, we consider here lattices with circular boundary which are well-approximated by translation invariant Majorana tight-binding models. The case of a generic boundary -- a candidate for future studies -- is different as there is an annular region with apparently disordered bonds but which still leave a finite fraction of bulk Majorana states at zero energy (see supplementary). In closing, we note that understanding the effects of including the EM gauge field\cite{2000-franz-uq, 2001-vafek-fk} as well as the effect of SC phase profiles that are not Abrikosov lattice-like are also challenging avenues of future research.

\begin{acknowledgments}
We thank Michael Stone, Eduardo Fradkin, Smitha Vishveshwara and Nadya Mason for insightful discussions. We would also like to acknowledge useful discussions with Eytan Grosfeld and Ady Stern about their past research on related topics. This research was supported by the Institute of Condensed Matter Theory at UIUC.
\end{acknowledgments}

\end{document}